\documentclass[10pt,letterpaper]{article}
\usepackage[top=0.85in,left=1.5in,footskip=0.75in]{geometry}

\usepackage{amsmath,amssymb}

\usepackage{changepage}

\usepackage[utf8x]{inputenc}

\usepackage{textcomp,marvosym}

\usepackage{cite}

\usepackage{nameref,hyperref}

\usepackage[right]{lineno}

\usepackage{microtype}
\DisableLigatures[f]{encoding = *, family = * }

\usepackage[table]{xcolor}

\usepackage{array}

\newcolumntype{+}{!{\vrule width 2pt}}

\newlength\savedwidth

\newcommand\thickhline{\noalign{\global\savedwidth\arrayrulewidth\global\arrayrulewidth 2pt}%
\hline
\noalign{\global\arrayrulewidth\savedwidth}}

\raggedright
\usepackage[aboveskip=1pt,labelfont=bf,labelsep=period,justification=raggedright,singlelinecheck=off]{caption}

\makeatletter
\renewcommand{\@biblabel}[1]{\quad#1.}
\makeatother

\usepackage{lastpage,fancyhdr,graphicx}
\usepackage{epstopdf}
\fancyheadoffset[L]{1.25in}
\fancyfootoffset[L]{1.25in}
\usepackage{graphicx}

\begin{document}
\vspace*{0.2in}

\begin{flushleft}
{\Large
\textbf\newline{Optimized protocol for DNA extraction from ancient skeletal remains using Chelex-100} 
}
\newline
\\
Wera M Schmerer
\\
\bigskip

\textbf{}Department of Biology, Chemistry and Forensic Science, Wolverhampton School of Science, University of Wolverhampton, Wolverhampton, UK
\\
\bigskip

w.schmerer@wlv.ac.uk

\end{flushleft}

\section*{Key words}
Degraded DNA, skeletal remains, DNA extraction, Chelex-100, inhibitors, optimization, STR-genotyping, human identification
  
\section*{Abstract}
PCR-based analysis of skeletonized human remains is a common aspect in both forensic human identification as well as Ancient DNA research. In this, both areas not merely utilize very similar methodology, but also share the same problems regarding quantity and quality of recovered DNA and presence of inhibitory substances in samples from excavated remains. To enable amplification based analysis of the remains, development of optimized DNA extraction procedures is thus a critical factor in both areas.
The study here presents an optimized protocol for DNA extraction from ancient skeletonized remains using Chelex-100, which proved to be effective in yielding amplifiable extracts from sample material excavated after centuries in a soil environment, which consequently have high inhibitor content and overall limited DNA preservation. Success of the optimization strategies utilized is shown in significantly improved amplification outcomes compared to the predecessor method.

\section*{Introduction}

Ancient DNA analysis of historical human remains explores similar questions, utilizing similar (and frequently the same) methodology as applied in the forensic human identification context, while doing so under extreme conditions with regards to DNA content, degree of degradation and presence of inhibitors. Consequently, improvements of methodology and procedures in one of these areas will inform the other, and vice versa.
\paragraph{}

When analyzing highly degraded or ancient DNA, the quality and quantity of DNA targets available for PCR amplification - and therefore the efficiency of the extraction process by means of which the genetic material is isolated - represents a crucial factor \cite{bib1} \cite{bib2} \cite{bib3}.  Consequently, the evaluation [e.g. see \cite{bib4}] and optimization of standard protocols \cite{bib2} \cite{bib3} \cite{bib5},  and the design of new protocols for the extraction of degraded DNA [e.g. \cite{bib6}, \cite{bib7}, \cite{bib8}, \cite{bib9}], are important means to improve the reliability of the analysis of degraded or ancient DNA.
\paragraph{}

Comparative studies including a variety of extraction procedures have shown that phenol/chloroform extraction protocols represent the most effective method to isolate amplifiable DNA [e.g. \cite{bib10}, \cite{bib4}], especially when extracting from hard tissues \cite{bib10} or even ancient human remains \cite{bib4}. However, the application of an organic extraction procedure might not always be possible, or researchers might be inclined to apply a less hazardous non-organic – but still very potent [e.g. \cite{bib11}, \cite{bib12}] – alternative DNA extraction protocol like the Chelex-100 method  \cite{bib13}. 
\paragraph{}

In context with the extraction of DNA from historical and ancient specimens, a further advantage of Chelex based DNA extraction is its applicability to minute samples of less than 1mg of bone or tooth powder, which significantly minimizes the damage to the analyzed specimen [14]. When analyzing samples containing minute quantities of DNA, an important advantage of Chelex based procedures is the comparatively limited risk of contamination with pristine modern DNA, due to the limited number of additions and transfers of reagents [cf. \cite{bib13}].
\paragraph{}

The disadvantage of standard Chelex protocols lies in their limited purification efficiency where samples containing PCR inhibitors are concerned: standard Chelex based extraction procedures may remove inhibitory substances to a certain extent \cite{bib15} \cite{bib16}. Depending on the concentration of the inhibitor present, additional purification of extracts may be required to allow for successful PCR amplification \cite{bib17}. In the case of low DNA content samples, the additional purification step can likewise be utilized as a concentration step to enrich DNA extracts \cite{bib18}.
Strategies for additional purification range from simple procedures, such as chromatography with Sephadex G-50 columns \cite{bib19} or ultrafiltration dialysis [e.g. \cite{bib20}], to a more complex re-extraction of extracts using silica column based commercial kits \cite{bib25}. A further method for purification of extracts is alcoholic precipitation of the DNA in the presence of sodium acetate \cite{bib21}. In the case of inhibitors like humic substances, which are frequently present in ancient specimen - especially those recovered from soil \cite{bib22} \cite{bib23}, replacement of the generally used ethanol \cite{bib21}  by isopropanol proved to be more efficient in removing inhibitory substances from DNA extracts \cite{bib7} . To support the precipitation of minute amounts of highly degraded DNA, as is usually encountered in archaeological or ancient skeletal material, silica can be added during the alcoholic precipitation \cite{bib2} \cite{bib24}.
\paragraph{}

Based upon a Chelex protocol published by Lassen \cite{bib25} for the extraction of DNA from ancient bone, a modified protocol was designed taking into account the findings published for successful optimization of a phenol/chloroform protocol for extraction of degraded DNA \cite{bib2} \cite{bib5}. The aim here was to improve this Chelex based method in terms of quantity and quality of extracted DNA. The optimized protocol (see supplementary material for full detail) was evaluated by comparison of the resulting extracts, to extracts of the same samples derived from the previously published protocol \cite{bib25}. 

\section*{Material and Methods}

The Chelex protocol published by Lassen \cite{bib25} consists of a decalcification of 0.3g bone powder in 0.45ml EDTA (0.5M, pH 8.3) for 24h at RT, followed by incubation with 0.3ml 5\% (w/v) Chelex solution and 10µl proteinase K (20mg/ml) at 56˚C for 45min, denaturation of the enzyme at 94˚C for 8min, subsequent concentration and purification applying the Wizard PCR Prep™ DNA Purification System (Promega) and final elution in 50µl sterile water. 
With reference to previously described findings \cite{bib2} \cite{bib5}, decalcification was extended to 48h at a constant temperature of 20˚C (incubator or shaking water bath) and digestion was carried out with an increased amount (100µl) proteinase K for an extended incubation time of 90min. Following the denaturation of the enzyme, a precipitation step in the presence of isopropanol (abs.), silica (e.g. Glasmilk™, Bio 101), and sodium acetate buffer (2M, pH 4.5) at pH 7.5 was added. The volume of the final extract is 50µl in sterile water.
For the comparison of both protocols a double set of extracts from each sample and protocol was prepared based on identical amounts of powdered bone prepared from seven specimens (historical skeletal remains).
\paragraph{}

Extracts were evaluated by PCR amplification of the two STR loci HUMVWA31/A \cite{bib26}  and HUMTH01 \cite{bib27}. Both loci were amplified in standardized PCR reactions using protocols optimized for reduced generation of shadow bands \cite{bib28} or stutter products \cite{bib29}, according to previously reported findings \cite{bib30} \cite{bib31}. Extracts derived from each protocol under comparison were amplified twice at each locus using a Mastercycler (Eppendorf).

For amplification of the locus HUMVWA, 5µl DNA extract (one tenth of the extract volume) was amplified in a total reaction volume of 50µl consisting of 16mM (NH\textsubscript{4})\textsubscript{2}SO\textsubscript{4}, 50mM Tris-HCl, 0.01\% Tween 20, 2mM MgCl\textsubscript{2}, 175µM of each dNTP, 6pmol (0.12µM) of each primer, 25µg/ml BSA, and 2U InViTAQ™ (InViTek). Cycling conditions consisted of an initial denaturation at 94˚C for 10min, followed by 60 cycles with 94˚C for 30sec, 50˚C for 1min, and 70˚C for 2min. 

For  HUMTH01, 5µl DNA extract (one tenth of the extract volume) from the bone material was again amplified in a total volume of 50µl consisting of 19.2mM (NH\textsubscript{4})\textsubscript{2}SO\textsubscript{4}, 60mM Tris-HCl, 0.012\% Tween 20, 2mM MgCl\textsubscript{2}, 200µM of each dNTP, 6pmol (0.12µM) of each primer, 25µg/ml BSA and 2U InViTAQ™ (InViTek). Cycling conditions consisted of an initial denaturation at 94˚C for 10min, followed by 55 cycles with 94˚C for 30sec, 53˚C for 1min, and 70˚C for 2min. 

Products of the amplification of these loci were processed applying fragment length analysis using a 310 Genetic Analyzer (Applied Biosystems) and the attached GeneScan Analysis 3.1 software (data not shown). Intensity of amplified products and amplification rates were evaluated by gel electrophoresis on 2.5\% standard ethidium bromide stained agarose gels. Fig. \ref{fig1} shows products of the amplification on locus HUMVWA from extracts of an exemplary subset of the samples included in this study.

\section*{Results and Discussion}

\begin{figure}[!h]
\includegraphics[width=5.5in]{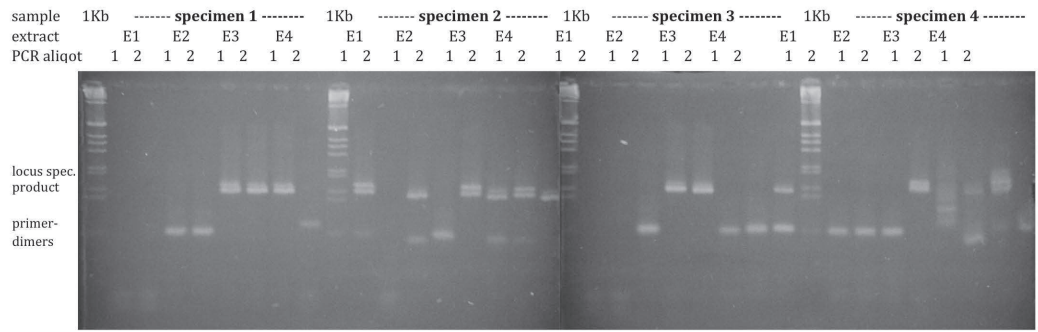}
\centering
\caption{{\bf Comparison of the standard (extracts E1 and E2) and optimized protocol (extracts E3 and E4) in amplifications of STR locus HUMVWA31/A.}
Agarose electrophoresis (2.5\% standard agarose gel) representing amplification products of a subset of the samples as extracted applying the two protocols to be compared. E1, E2: standard Chelex Protocol (Lassen \cite{bib25}). E3, E4: optimized Chelex protocol. 1Kb: 1 Kb DNA-ladder (Life Technologies)
Amplificates derived from optimized extracts show a comparably higher frequency of specific products at this locus, indicating a higher quality of isolated DNA compared to that found in basic extracts of the same material, since the amplifiability of the locus HUMVWA31/A is influenced by the quality of the target amplified \cite{bib32}}.
\label{fig1}
\end{figure}

In the comparison of both protocols, extracts prepared according to the new protocol (extracts E3 and E4) resulted in a significantly higher success rate in the amplification of locus HUMVWA, compared to those derived from the original protocol (extracts E1 and E2).  Locus specific product was seen in 70.8\% of amplificates using the new protocol, whereas only 33.3\% of amplificates using the original protocol showed locus specific product (see Fig. \ref{fig1}). Amplifications at the locus HUMTH01 resulted in equal success rates for both protocols.
\paragraph{}

In general, the amplification success of HUMVWA is reduced with decreasing quality (with regards to structural damage) of the target amplified, while HUMTH01 amplifications are rendered less efficient by decrease in DNA quantity (number of amplifiable copies of target DNA) \cite{bib32}. Hence, the results obtained indicate that both methods may yield similar quantities of DNA, with but a comparably higher target quality in the case of the optimized protocol.
In addition to the above findings, 20.8\% of the amplifications from standard extracts 1 and 2 at locus HUMVWA were inhibited (easily recognizable by absence of locus-specific product as well as primer-dimers), while no inhibition occurred in case of the new protocol preparations (cf. Fig. \ref{fig1}). Amplifications at the locus HUMTH01 again resulted in similar rates of inhibition for both extraction protocols. Keeping in mind that the amplification of HUMTH01 is less prone to inhibition by co-extracted impurities than that of HUMVWA \cite{bib32}], these results indicate a comparably higher efficiency of the new protocol in the removal of inhibiting agents due to the addition of the precipitation step in the presence of isopropanol [cf.\cite{bib8}].
\paragraph{}

The protocol presented here is still effective without addition of silica during the precipitation step, even when extracting trace amounts of DNA, as successfully demonstrated in a number of student projects supervised by the author:
Although originally developed for the extraction from ancient bone, the protocol can likewise be used successfully for the extraction of DNA from blood, dried blood spots and saliva [36], as well as trace DNA material like hair \cite{bib36} \cite{bib37} finger nail clippings, skin, cigarette ends with or without filter \cite{bib37} and even fingerprint residues \cite{bib38}. 
In tests with known concentration of polymerase-inhibitor in the extracted sample (humic acid added to saliva), this modified Chelex protocol was more efficient in removing the inhibitor than phenol-chloroform extraction \cite{bib39}, which concurs with previously published findings \cite{bib40} \cite{bib41} \cite{bib42} \cite{bib43} \cite{bib17} \cite{bib18}. Applying this protocol, the inhibitor was successfully removed even when present in high concentrations \cite{bib39}.
These studies \cite{bib36} \cite{bib37} \cite{bib38} \cite{bib39} likewise showed the protocol to be stable and easy to use, even in the hand of relatively inexperienced (student) researchers.

\section*{Supplementary Material}

\paragraph*{Optimized Protocol}
A detailed version of the optimized protocol can be found as addendum to this article, following the reference section.

\section*{Acknowledgments}
The author would like to thank Dr Terence J Bartlett (University of Wolverhampton) for helpful discussions and pertinent comments.

\section*{Supplementary Material}

\paragraph*{Optimized Protocol}
\paragraph{I. Chelex 100-based DNA extraction}

\begin{enumerate}
	\item{When investigating ancient human skeletons, the sampling of long bones – preferably the mid shaft region – is recommendable. Because of their compactness, these skeletal elements show a relatively high probability of successful DNA recovery compared to less dense parts of the skeleton. A sample of 1x1cm (ca. 1g) is separated for processing as follows:}
	\item{To prevent co-processing of possible adhering contaminations, exposed surfaces of the bone sample are quantitatively removed by the use of a sterile scalpel. Subsequently the material is exposed to UV light for 15min each of the previous surfaces (periosteum and medullary cavity).}
	\item{Samples are ground to a fine powder using a mixer mill (MM2000, Retsch) or an agate mortar and pestle, according to the consistency of the material.}
	\item{0.3g bone powder is mixed with 1,5ml EDTA-solution (0.5M, pH 8.3) in a 2ml reaction tube, vortexed vigorously (e.g. Vortex Genie 2, VWRbrand), and incubated at constant rotation or agitation respectively for 48h at a constant temperature of 20˚C (incubator or shaking water bath). 
	Depending on the degree of DNA degradation respectively state of DNA preservation to be expected in the material at hand, the parameters of this step can be adapted to the following to optimize DNA yield of the extracts [cf. \cite{bib2} \cite{bib5}]:}
\paragraph{}	

% Place table
\begin{table}[htp]
\centering
\caption{\bf Parameters for optimized decalcification of bone based on the expected preservation of the contained DNA.}
\begin{center}
\begin{tabular}{|l+l|l|l|l|l|l|l|}
\hline
\multicolumn{1}{|l|}{\bf Degree of DNA degradation} & \multicolumn{1}{|l|}{\bf Incubation time} & \multicolumn{1}{|l|}{\bf Temperature}\\ \thickhline
high & 96-120h & 20˚C\\ \hline
intermediate & 48h &(20)-30˚C\\ \hline
low & 24h & (20)-30˚C\\ \hline
consensus (DNA preservation n.d.) & 96h & 20˚C\\ \hline
\end{tabular}
\end{center}
\label{default}
\end{table}

	\item{The remaining bone powder is pelleted by centrifugation for 5min at 6000rpm (desktop centrifuge  5415C, Eppendorf).}
	\item{The supernatant (ca. 1300µl) is transferred to a 5ml tube (Polypropylene round-bottom tube, Falcon).}
	\item{1300µl Chelex-100-solution (5\% in sterile Water, e.g. Ampuwa\textsuperscript{®}, Fresenius or 18 Megohm Water, Sigma) and 500µl proteinase K-solution (20mg/ml, e.g. Qiagen) are added.}
	\item{The mixture is vortexed briefly (5-10sec, Vortex Genie 2, VWRbrand) and incubated at 56˚C and constant shaking at 300rpm (Thermomixer, Eppendorf) for a duration appropriate for the DNA degradation expected in the material to be analyzed \cite{bib2} \cite{bib5}:}
	
% Place table
\begin{table}[htp]
\centering
\caption{\bf Parameters for optimized proteinase K digestion from decalcified bone based on the expected preservation of the contained DNA.}
\begin{center}
\begin{tabular}{|l+l|l|l|l|l|l|l|}
\hline
\multicolumn{1}{|l|}{\bf Degree of DNA degradation} & \multicolumn{1}{|l|}{\bf Incubation time}\\ \thickhline
high & (60)-90min\\ \hline
intermediate & 90min\\ \hline
low & 60min\\ \hline
consensus (DNA preservation n.d.) & 90min\\ \hline
\end{tabular}
\end{center}
\label{default}
\end{table}	
	
	\item{The mix is vortexed again briefly (5-10sec) and incubated at 95˚C for 8min (Thermomixer, Eppendorf) to denature and deactivate the proteinase.}
	\item{The mixture is left to cool down slowly to room temperature (Thermomixer, Eppendorf) and the aqueous portion is separated from the Chelex resin by centrifugation for 6min at 4000rpm (Centrifuge 5804, Eppendorf).}
\paragraph{II. Alcoholic precipitation in the presence of silica}
	\item{The aqueous supernatant is transferred to a 14ml tube (Polypropylene Round-Bottom Tube, Falcon) with addition of 3250µl Isopropanol (abs., RT), 60-120µl sodium acetate buffer (2M, pH 4.5) and 5µl silica solution (Glasmilk™, Bio 101).}
	\item{Prior to addition of the silica, the pH of the solution should be evaluated (e.g. PH-Indicator Strips, pH 6.5-10, Merck). If necessary, the pH should be adjusted to a value of 7.5 by adding further sodium acetate buffer to ensure optimal adsorptive binding of DNA to silica [cf. \cite{bib2} \cite{bib5}].}
	\item{Precipitation is carried out for 30min at RT.}
	\item{Subsequently the precipitate is separated by centrifugation for 2min at 4000rpm (Centrifuge 5804, Eppendorf) and the supernatant discarded by careful decanting.}
	\item{The DNA-silica pellet is washed with 500µl EtOH (abs.), then the alcohol removed by centrifugation for 2min at 4000rpm (Centrifuge 5804, Eppendorf) and the pellet left to air dry for ca. 30min at RT.}
	\item{The DNA is eluted in 50µl sterile water (e.g. Ampuwa\textsuperscript{®}, Fresenius or 18 Megohm Water, Sigma) for 5min at 50˚C and constant shaking at 300rpm (Termomixer, Eppendorf) and the eluate transferred to a 2ml tube (Safelock, Eppendorf). To ensure optimal stability of the extracted DNA, storage of the extract with silica at -20˚C is recommended \cite{bib33}.}

\end{enumerate}

\paragraph{}
PCR inhibition due to co-amplified polymerase inhibiting substances [e.g. \cite{bib22}], or brownish color of bone powder or the resulting extract indicate the presence of inhibitors like humic acids \cite{bib23} \cite{bib34}, as frequently present when amplifying DNA extracted from historical or ancient specimen.  In these cases an additional cleaning of the extract would be indicated. For this purpose the application of e.g. ultrafiltration dialysis [e.g. \cite{bib20}] or the Wizard PCR Prep™ DNA Purification System (Promega) \cite{bib35} following a modified protocol \cite{bib5} could be utilized.

\end{document}